\documentclass[twocolumn]{aastex631}

\newcommand{\vdag}{(v)^\dagger}

\usepackage{array, tabularx, xcolor}

\begin{document}

\title{Clumpy, dense gas in the outflow of NGC 1266}

\author[0000-0003-3191-9039]{Justin Atsushi Otter}
\email{jotter2@jhu.edu}
\affiliation{William H. Miller III Department of Physics and Astronomy, Johns Hopkins University, Baltimore, MD 21218, USA}

\author[0000-0002-4261-2326]{Katherine Alatalo}
\affiliation{Space Telescope Science Institute, 3700 San Martin Drive, Baltimore, MD 21218, USA}
\affiliation{William H. Miller III Department of Physics and Astronomy, Johns Hopkins University, Baltimore, MD 21218, USA}

\author[0000-0001-7883-8434]{Kate Rowlands}
\affiliation{AURA for ESA, Space Telescope Science Institute, 3700 San Martin Drive, Baltimore, MD 21218, USA}
\affiliation{William H. Miller III Department of Physics and Astronomy, Johns Hopkins University, Baltimore, MD 21218, USA}

\author[0000-0002-9471-8499]{Pallavi Patil}
\affiliation{William H. Miller III Department of Physics and Astronomy, Johns Hopkins University, Baltimore, MD 21218, USA}

\author[0009-0004-0844-0657]{Maya Skarbinski}
\affiliation{William H. Miller III Department of Physics and Astronomy, Johns Hopkins University, Baltimore, MD 21218, USA}

\author[0009-0005-3525-1904]{Lauren Dysarz}
\affiliation{William H. Miller III Department of Physics and Astronomy, Johns Hopkins University, Baltimore, MD 21218, USA}

\author[0000-0002-3032-1783]{Mark Lacy}
\affiliation{National Radio Astronomy Observatory, Charlottesville, VA 22903, USA}

\author[0000-0002-9165-8080]{Mar\'ia J. Jim\'enez-Donaire}
\affiliation{AURA for ESA, Space Telescope Science Institute, 3700 San Martin Drive, Baltimore, MD 21218, USA}
\affiliation{Observatorio Astron\'omico Nacional (IGN), C/ Alfonso XII, 3, 28014 Madrid, Spain}

\author{Susanne Aalto}
\affiliation{Department of Space, Earth and Environment, Chalmers University of Technology, 412 96 Gothenburg, Sweden}

\author[0000-0003-4932-9379]{Timothy A. Davis}
\affiliation{Cardiff Hub for Astrophysics Research \& Technology, School of Physics \& Astronomy, Cardiff University, Queens Buildings, Cardiff CF24 3AA, UK}

\author[0009-0007-2501-3931]{Antoniu Fodor}
\affiliation{Department of Physics \& Astronomy and Ritter Astrophysical Research Center, University of Toledo, Toledo, OH 43606, USA}

\author[0000-0002-4235-7337]{K. Decker French}
\affiliation{Department of Astronomy, University of Illinois, 1002 W. Green Street, Urbana, IL 61801, USA}

\author[0000-0002-6824-6627]{Nanase Harada}
\affiliation{National Astronomical Observatory of Japan, 2-21-1 Osawa, Mitaka, Tokyo 181-8588, Japan}
\affiliation{Astronomical Science Program, Graduate Institute for Advanced Studies, SOKENDAI, 2-21-1 Osawa, Mitaka, Tokyo 181-1855, Japan}

\author[0000-0001-6670-6370]{Timothy Heckman}
\affiliation{William H. Miller III Department of Physics and Astronomy, Johns Hopkins University, Baltimore, MD 21218, USA}
\affiliation{School of Earth and Space Exploration, Arizona State University, }

\author[0009-0005-0757-8547]{Ryo Kishikawa}
\affiliation{Department of Astronomy, The University of Tokyo, 7-3-1 Hongo, Bunkyo, Tokyo 113-0033, Japan}
\affiliation{National Astronomical Observatory of Japan, 2-21-1 Osawa, Mitaka, Tokyo 181-8588, Japan}

\author[0000-0002-2644-0077]{Sebastian Lopez}
\affiliation{Department of Astronomy, The Ohio State University, 140 W. 18th Ave., Columbus, OH 43210, USA}
\affil{Center for Cosmology and AstroParticle Physics, The Ohio State University, 191 W. Woodruff Ave., Columbus, OH 43210, USA}

\author[0000-0002-0696-6952]{Yuanze Luo}
\affiliation{Department of Physics and Astronomy and George P. and Cynthia Woods Mitchell Institute for Fundamental Physics and Astronomy,Texas A\&M University, 4242 TAMU, College Station, TX 77843-4242, US}

\author[0000-0001-9281-2919]{Sergio Martin}
\affiliation{European Southern Observatory, Alonso de Córdova, 3107, Vitacura, Santiago 763-0355, Chile}
\affiliation{Joint ALMA Observatory, Alonso de Córdova, 3107, Vitacura, Santiago 763-0355, Chile}

\author[0000-0001-7421-2944]{Anne M. Medling}
\affiliation{Department of Physics \& Astronomy and Ritter Astrophysical Research Center, University of Toledo, Toledo, OH 43606, USA}

\author[0000-0003-1991-370X]{Kristina Nyland}
\affiliation{U.S. Naval Research Laboratory, 4555 Overlook Ave SW, Washington, DC 20375, USA}

\author[0000-0003-4030-3455]{Andreea Petric}
\affiliation{Space Telescope Science Institute, 3700 San Martin Drive, Baltimore, MD 21218, USA}

\author[0000-0002-4430-8846]{Namrata Roy}
\affiliation{William H. Miller III Department of Physics and Astronomy, Johns Hopkins University, Baltimore, MD 21218, USA}

\author{Mamiko Sato}
\affiliation{Department of Space, Earth and Environment, Chalmers University of Technology, 412 96 Gothenburg, Sweden}

\author[0000-0001-6245-5121]{Elizaveta Sazonova}
\affiliation{Department of Physics and Astronomy, University of Waterloo, 200 University Avenue West, Waterloo, ON N2L 3G1, Canada}

\author[0000-0003-2599-7524]{Adam Smercina}
\affiliation{Space Telescope Science Institute, 3700 San Martin Drive, Baltimore, MD 21218, USA}

\author[0000-0002-6582-4946]{Akshat Tripathi}
\affiliation{Department of Astronomy, University of Illinois, 1002 W. Green Street, Urbana, IL 61801, USA}

\begin{abstract}
Outflows are one of the most spectacular mechanisms through which active galactic nuclei (AGN) impact their host galaxy, though the role of AGN-driven outflows in global star formation regulation across the galaxy population is unclear.
NGC 1266 is an excellent case study for investigating the outflows and star formation quenching because it is a nearby (D$\sim$30 Mpc) AGN host galaxy with an outflow driving shocks through the interstellar medium (ISM) and has recently quenched its star formation outside the nucleus.
While previous works have studied the molecular outflow from its CO emission, to fully characterize the impact the outflow has on the ISM observations probing the dense, cold gas are necessary.
Our ALMA cycle 0 observations do not detect a molecular outflow  in $^{13}$CO(2-1) and yield a lower limit $^{12}$CO/$^{13}$CO $\geq$ 250, suggesting a highly optically thin CO outflow with low $^{13}$CO abundance.
In contrast, we detect substantial HCN(1-0) emission in the outflow, with an HCN(1-0)/$^{12}$CO(1-0) ratio of 0.09, consistent with global measurements of many star-forming galaxies and Luminous InfraRed Galaxies (LIRGs).
We conclude that the CO emission traces a diffuse component of the molecular gas with a low optical depth, whereas the HCN(1-0) traces dense clumps of gas entrained in the outflow.
We measure an upper limit molecular outflow rate of $<85$ M$_\odot$/yr.
Assuming the ongoing nuclear star formation and outflow continue at the same rates, NGC 1266 will deplete its gas reservoirs in 450 Myr or longer, indicating that relatively low-level AGN feedback is capable of gradually expelling the molecular gas reservoir after a rapid quenching event.
\end{abstract}


\section{Introduction} \label{sec:intro}

Through cosmic time, galaxies evolve from star-forming to quiescent, transforming many of their observable properties including their morphology, kinematics, and color \citep{bell07, bell12, ilbert13}.
The global star formation rate of a galaxy is strongly linked to the aforementioned directly observable properties, as the distribution of these properties among the galaxy population is bimodal with two primary galaxy archetypes: blue, star-forming spiral galaxies, and red, quiescent elliptical galaxies \citep{kauffmann03c, baldry04, jin14a}.
The global star formation rate of a galaxy can be impacted by a variety of physical processes, including stellar feedback \citep[e.g.][]{veilleux05,veilleux20}, gas stabilization from the galaxy's stellar morphology \citep{martig09a, gensior20}, gas disruption from stellar bars \citep[e.g.][]{scaloni24}, outflows and winds from stellar and active galactic nuclei (AGN) feedback \citep[e.g.][]{feruglio10, lambrides19a}, galaxy interactions and mergers \citep[e.g.][]{woods07, ellison08, scudder12, ellison18}, and gas accretion from the inter-galactic medium \citep{keres05}.
While each of these mechanisms shape the conditions that regulate and quench star formation in galaxies, much work is needed to understand both the relative impacts of galaxy-scale processes on star formation, as well as the physics coupling these mechanisms to local cloud properties where stars form.

Though observational evidence and simulations demonstrate that feedback from a central supermassive black hole is necessary to regulate star formation through cosmic time \citep{bower06, croton06, terrazas17, terrazas20, dave19}, the physical mechanisms through which an AGN can impact the host galaxy's ISM are poorly understood.
AGN feedback is typically classified as either kinetic (radio mode) or radiative (quasar mode), where the former is powered by relativistic radio jets at low accretion rates from the central SMBH, and the latter by radiative pressure at high accretion rates \citep{fabian12, heckman14, heckman23}.
In simulations, AGN feedback is critical in reproducing observed properties of the Universe, such as the stellar mass distribution function and the low overall current star formation rate \citep[e.g.][]{kauffmann99, dimatteo05, hopkins06b, vogelsberger14, schaye15}.
Further, observational evidence shows that powerful AGN in massive central galaxies can expel their gas reserves and heat the surrounding intergalactic medium, ultimately quenching their star formation activity by removing the gas and preventing the surrounding gas from re-accreting onto the host galaxy \citep[e.g.][]{mcnamara00, fabian00}.
While AGN-driven gas expulsion appears to be an important quenching pathway for massive galaxies, less massive galaxies with weaker AGN activity are often unable to totally expel their molecular gas reservoirs \citep{fluetsch19, luo22}.
Large scale simulations further are unable to resolve the detailed physical conditions of outflows, making it difficult to determine both how much gas leaves a galaxy and whether outflows contain dense gas \citep{crain23}.

Furthermore, studies of the molecular gas content of recently quenched galaxies demonstrate that galaxies can quench their star formation \textit{without} gas removal.
Post-starburst galaxies (PSBs) provide an excellent laboratory to study star formation quenching, as their dominant A-type stellar populations and lack of ongoing star formation indicate that they have rapidly quenched their star formation in the past $\sim$1 Gyr.
Many PSBs retain their molecular gas reservoirs after quenching their star formation \citep{french15a, rowlands15a, alatalo16d}; though it is unclear why there is little ongoing star formation in the remaining molecular gas.
One possibility is that outflows incapable of ejecting the molecular gas from the galaxy may make the remaining molecular gas inhospitable to star formation by kinematically heating the gas \citep{alatalo11a, alatalo15c, luo22, smercina22b}.
However, to build a full picture of star formation suppression in these galaxies, it is necessary to study the denser, cold phases of the molecular gas that are most proximal to star formation \citep[e.g.][]{gao04b, wu05, wu10b, lada10, lada12}.

NGC 1266 is a nearby (30 Mpc, $z=0.007$) quenching early-type galaxy with log$M_*/M_\odot\sim10.4$, where turbulence injected by outflows has been suggested as a mechanism driving star formation suppression \citep{alatalo11a, alatalo15c}.
$^{12}$CO observations of this system show a centrally concentrated, high surface density molecular gas reservoir with bipolar outflows \citep{alatalo11a}.
Any remaining star formation in this galaxy is restricted to the nucleus, and thus the bulk of the molecular gas that is beyond the nucleus has highly suppressed star formation.
The nuclear star formation rate of 0.7 M$_\odot$/yr is insufficient to drive previously measured molecular outflow rates of 10-100 M$_\odot$/yr \citep{alatalo15c, otter24}, indicating that the outflow is likely driven by an AGN.
X-ray and Very Long Baseline Array (VLBA) milliarcsecond resolution radio imaging provide strong evidence that NGC 1266 hosts an AGN \citep{alatalo11a, nyland13}, though multi-wavelength measurements of the bolometric luminosity of the AGN yield differing results.
IR photometry from \citet{alatalo15c} indicate a bolometric luminosity of $L_{bol}\sim10^{43}$ erg/s, optical SED fitting from \citet{chen23} and X-ray observations from Lanz et. al. (in prep) yield a value closer to 10$^{42}$ erg/s.

These outflows appear to drive low-velocity C-type shocks through the molecular gas of NGC 1266, as traced by ro-vibrational H$_2$, H$_2$O, and high-J CO emission lines \citep{pellegrini13a, otter24}. 
The presence of shocks in the ISM provides evidence that the outflow is interacting with the ambient molecular gas in the galaxy, and thus the outflow could plausibly inject turbulence into the molecular gas reservoir and contribute to star formation suppression.
However, dense molecular gas tracers are needed to understand the physics underlying the energy transfer between the outflow and surrounding ISM and its impact on star formation.
HCN has long been a commonly used tracer of dense molecular gas because it is relatively bright, and typically traces denser gas than CO \citep[around $10^3-10^4$ cm$^{-3}$ for HCN and $\sim10^2$ cm$^{-3}$ for CO][]{shirley15, jones23}.

In this work we present archival ALMA observations of NGC 1266, focusing on the HCN(1-0) and $^{13}$CO(2-1) emission lines.
In Section~\ref{sec:obs} we describe the observations and reduction procedure.
In Section~\ref{sec:results} we present our results and in Section~\ref{sec:discuss} discuss their implications.
Finally, we present our conclusions in Section~\ref{sec:conclusions}.
Throughout this work, we adopt a Hubble constant of $H_0 = 70$km s$^{-1}$ Mpc$^{-1}$, $\Omega_m$ = 0.3, and $\Omega_\Lambda$ = 0.7 for computing distances and spatial scales.

\section{Observations} \label{sec:obs}
ALMA Band 3, 6, and 7 observations of NGC 1266 were obtained with the 12 m array in Cycle 0 in the extended configuration as part of program 2011.0.00511.S (P.I. Alatalo).
In total, 11 spectral windows were observed, though in this work we focus on the band 3 and 6 spectral windows covering HCN(1-0) and $^{13}$CO(2-1).
Specific dates, on-source times, and configuration information for all observations are reported in Table~\ref{tab:obs}.
All observations were centered on the coordinates of the optical center (03:16:00.76, -02:25:38.40).
These data were calibrated with the observatory-provided ALMA pipeline using the Common Astronomy Software Applications package (\texttt{CASA}) version 3.4, as newer versions of \texttt{CASA} have compatibility issues with Cycle 0 observations \citep{mcmullin07a}.

We image the pipeline calibrated data with \texttt{CASA} version 6.6.4 and the task \texttt{tclean}.
The measurement sets were continuum subtracted in $u,v$ space with the \texttt{CASA} task \textit{uvcontsub} using line-free channels of each spectral window.
We apply Briggs weighting with a robust parameter of 2 (close to natural weighting) to maximize the signal to noise ratio of the integrated spectra.

For the HCN(1-0) and $^{13}$CO(2-1) observations, we additionally apply a taper in $u,v$ space to the weighting, further increasing the beam size and maximizing the signal to noise.
The full-width at half-maximum (FWHM) of the applied taper is 2\arcsec\, and 3\arcsec\, for the $^{13}$CO(2-1) and HCN(1-0) observations, respectively.
Finally, we remove baselines less than 30 m (22 k$\lambda$) for the $^{13}$CO(2-1) observations to remove large scale artifacts in the imaging.
The final reduced cubes for the HCN(1-0) and $^{13}$CO(2-1) have synthesized beam sizes of (3.9\arcsec$\times$3.3\arcsec) and (2.0\arcsec$\times$1.9\arcsec) and 1-sigma rms sensitivities of 2.1e-3 Jy/beam and 1.4e-3 Jy/beam respectively, both measured over 11 km/s channel widths.
The final spectral channel widths of the cubes are 3.23 MHz and 8.09 MHz (corresponding to 11 km/s for both).

For the remaining spectral windows, we image the measurement sets with Briggs weighting with a robust parameter of 0.5 in case any of the line emission is resolved.
We detect a number of molecular lines that we present in Appendix~\ref{app:lines}.

\begin{table*}[]
    \centering
    \begin{tabular}{c|c|c|c|c|c|c|c}
     Band & Date & On source & Angular & Continuum & 5th percentile & 80th percentile & Antennas \\ 
      & & time (s) & resolution ($\arcsec$) & rms (mJy/beam) & baseline (m) & baseline (m) &  \\ \hline
     Band 7 & Jun. 18th, 2012$^a$ & 4536.0 & 0.44 & 0.107 & 47.93 m & 233.34 m & 26 \\
     Band 7 & Aug. 24th, 2012 & 2721.6 & 0.55 & 0.067 & 39.14 m & 214.52 m & 25 \\
     Band 6 & Aug. 26th, 2012 & 2237.8 & 0.59 & 0.041 & 44.16 m & 235.33 m & 25 \\
     Band 3$^{*}$ & Aug. 27th, 2012 & 725.8 & 1.91 & 0.071 &  39.6 m & 215.05 m & 25 \\
     Band 6$^{\dagger}$ & Oct. 22nd, 2012 & 2147.0 & 0.78 & 0.050 & 38.11 m & 208.77 m & 23
    \end{tabular}
    \caption{Observations of NGC 1266 from ALMA Cycle 0. ($^*$) observation of HCN(1-0). ($^\dagger$) observation of $^{13}$CO(2-1). $^a$ This observation was split across two observing dates, June 18th, 2012 and July 27th, 2012.}
    \label{tab:obs}
\end{table*}

\section{Results and Analysis} \label{sec:results}

\subsection{Molecular gas extent} \label{sec:extent}

\begin{figure*}
    \centering
    \includegraphics[width=0.95\linewidth]{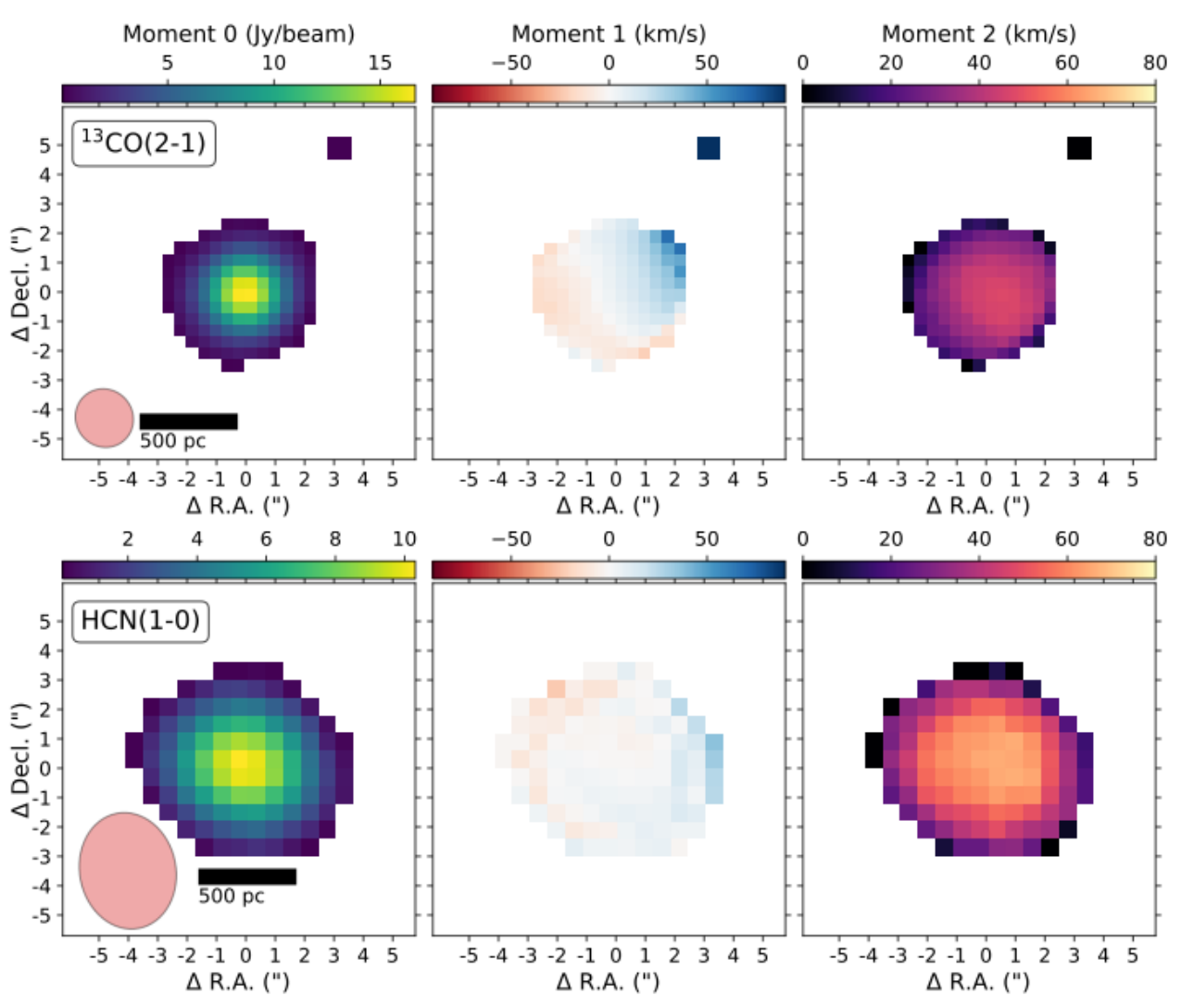}
    \caption{Moment 0, 1, and 2 maps of the ALMA Cycle 0 observations of NGC 1266. Only spaxels with S/N $\geq$ 5 in the moment 0 map are shown. The top row shows the moment maps for $^{13}$CO(2-1), and the bottom row shows the maps for HCN(1-0). The axes are in units of arcseconds offset from the center of the galaxy. The red ellipse in the lower left corner of the moment 0 maps show the ALMA beam size. All panels are centered on the coordinates (3:16:00.75, -2:25:38.70)}
    \label{fig:mom_maps}
\end{figure*}

In Figure~\ref{fig:mom_maps}, we plot the moment 0, 1, and 2 maps for $^{13}$CO(2-1) and HCN(1-0).
We generate these moment maps by extracting spectral slabs from $\pm$ 500 km/s of the observed line wavelength, masking all points with flux less than 5 times the rms noise (measured in a line-free region of the cube).
Finally, we use the \texttt{spectral-cube} python package \citep{ginsburg14} to create moment maps from these masked slabs.
The $^{13}$CO(2-1) emission is spatially resolved, spanning a region of approximately 4\arcsec\, (600 pc).
The $^{13}$CO(2-1) velocity field does not show a clear axis of rotation.
The HCN(1-0) emission is unresolved with our 3.9\arcsec\, (560 pc) beam.

We compare the extent of the HCN(1-0), $^{13}$CO(2-1), and CO(2-1) emission from \citet{alatalo11a} in Figure~\ref{fig:extent}.
The $^{13}$CO(2-1) emitting region is similar to the central CO(2-1) emission, though there is some extended CO(2-1) emission in the southwest, likely tracing outflowing gas \citep{alatalo11a}, that is undetected in our $^{13}$CO(2-1) observations.
For HCN(1-0), the physical beam size of 560 pc is a conservative upper limit on the emitting region.

\begin{figure}
    \centering
    \includegraphics[width=\linewidth]{HCN_CO_extent.pdf}
    \caption{Molecular gas extent in multiple gas tracers. The colormap shows the CO(2-1) emission, the teal contours show the $^{13}$CO(2-1) emission, and the green contours show the unresolved HCN(1-0) emission. The contour levels correspond to 15, 30, 60, and 90\% of the maximum flux. The green, red, and teal ellipses show the beam size for the HCN(1-0), CO(2-1), and $^{13}$CO(2-1) observations respectively.}
    \label{fig:extent}
\end{figure}

\subsection{Gaussian Fitting} \label{sec:fitting}

We extract our HCN(1-0) and $^{13}$CO(2-1) spectra from the continuum-subtracted cubes in a circular aperture with a radius of 5\arcsec.
We fit the spectra with a non-linear least squares method, specifically the trust region reflective fitting algorithm as implemented in \texttt{scipy}.
We perform two fits on each spectrum: one fit with two freely varying Gaussian components (labeled as fit A), and another two component Gaussian fit but with one component constrained to have $\sigma\sim145$ km/s (fit B), and for the peak velocities of the components to be within 25 km/s.
This linewidth was chosen to match the best-fitting broad component linewidth for CO(2-1) in \citet{alatalo11a}, and was also used for two component fits of CO(1-0) and CO(3-2) in that work.
In the $^{12}$CO and HCN spectra, the two kinematic components have peak velocities within 20 km/s, so we require that the components have similar peak velocities for our constrained fit.

\begin{figure*}[h]
    \centering
    \includegraphics[width=\linewidth]{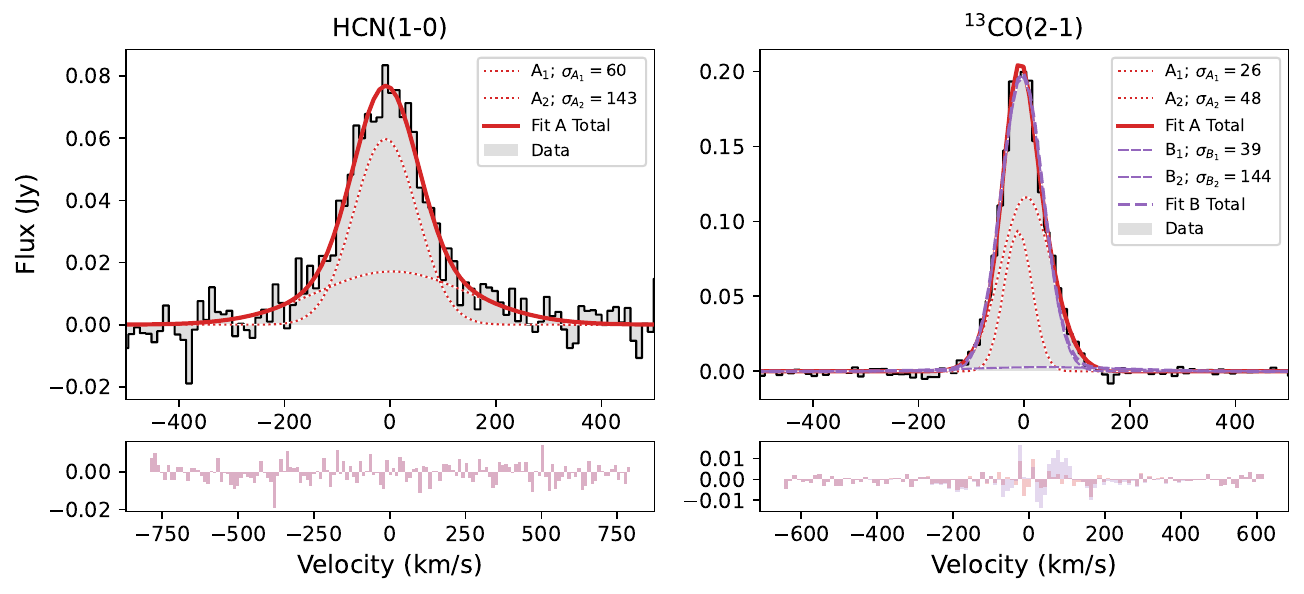}
    \caption{ALMA HCN(1-0) and $^{13}$CO(2-1) spectra of NGC 1266. Left: HCN(1-0) spectrum in gray bars. The red solid line is the unconstrained two Gaussian fit of the spectrum (fit A). The two red dotted lines show the narrow and wide Gaussian components. Below, the colored bars show the residuals of the plotted fits. Right: all symbols are the same as the left, but with $^{13}$CO(2-1). The thick purple dashed line is the two Gaussian fit of the spectrum with the broad Gaussian component having a fixed width of $\sigma$ = 145 km/s (fit B). The two thinner purple dashed lines show the narrow and broad Gaussian components of this fit. The zero velocity is set to a redshift of 0.0072.} Fit parameters are given in Table~\ref{tab:fit_params}. We detect a narrow and wide component in the HCN(1-0) spectrum, while the $^{13}$CO(2-1) spectrum is consistent with a single component.
    \label{fig:spectra}
\end{figure*}

\begin{figure}[h]
    \centering
    \includegraphics[width=\linewidth]{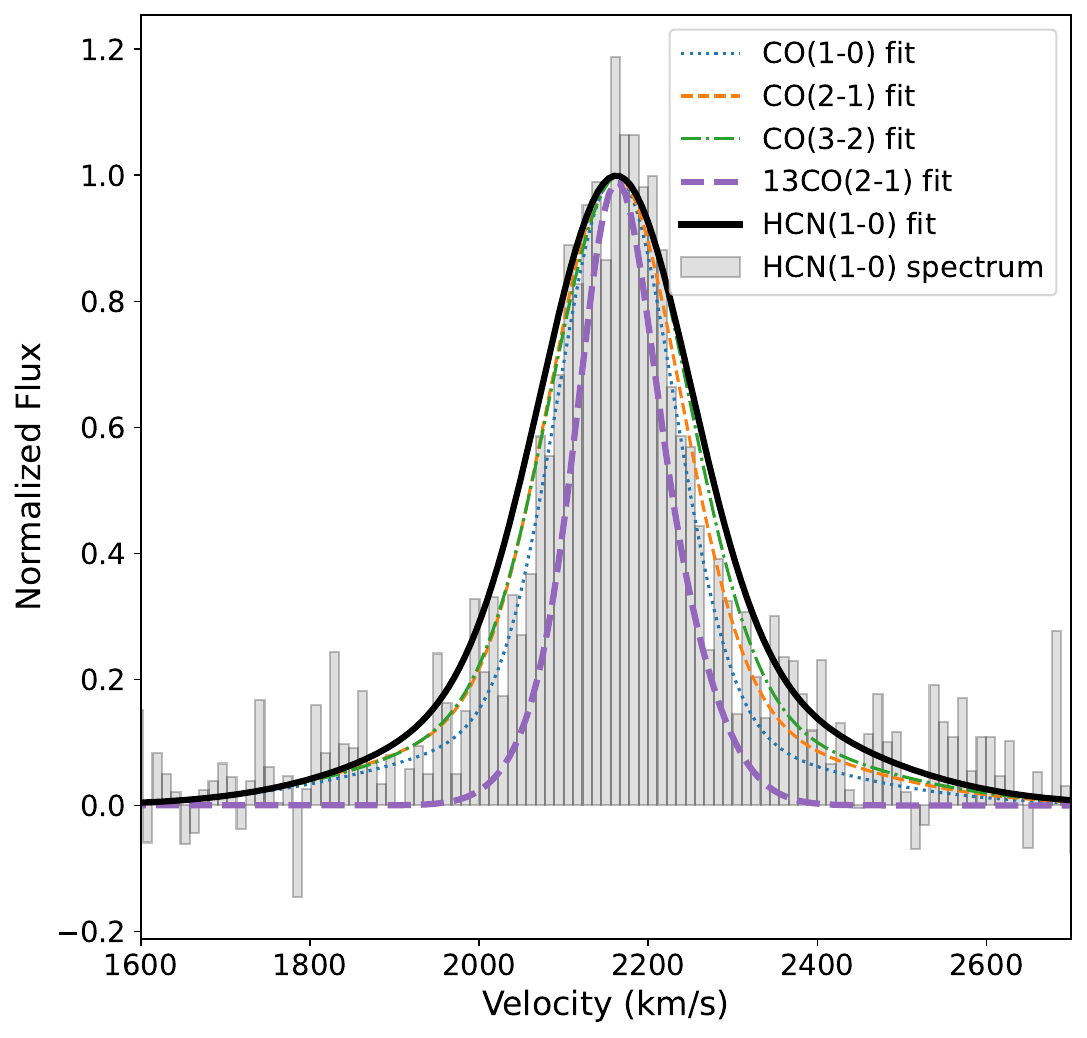}
    \caption{Fitted two component Gaussian line profiles of various molecular lines in NGC 1266. Three $^{12}$CO transitions from \citet{alatalo11a} are plotted: $^{12}$CO(1-0) in a dotted blue line, $^{12}$CO(2-1) in a dashed orange line, and $^{12}$CO(3-2) in a dot-dashed green line. From this work, we plot the profile of HCN(1-0) in the thick solid black line, and $^{13}$CO(2-1) in a thick dashed purple line. All profiles have been normalized for comparison. The gray bars show the observed HCN(1-0) spectrum. The $^{12}$CO (from \citealt{alatalo11a}) and the HCN(1-0) line profiles are similar, while the $^{13}$CO line profile is more narrow (this work).
    }
    \label{fig:line_profiles}
\end{figure}

In Figure~\ref{fig:spectra}, we plot the HCN(1-0) and $^{13}$CO(2-1) spectra and the constrained and unconstrained fits.
For HCN(1-0), the unconstrained (fit A) and constrained (fit B) fits agree within fitting uncertainties so we only plot the unconstrained fit.
In contrast, for $^{13}$CO(2-1), the constrained fit (in purple) poorly matches the spectrum, where the fitted amplitude for the required wide component is negligible, while the unconstrained fit favors two Gaussian components both with $\sigma < 50$ km/s.
We compare our unconstrained fitted line profiles with the line profiles of the original Gaussian fits for CO(1-0), CO(2-1), and CO(3-2) from \citet{alatalo11a} in Figure~\ref{fig:line_profiles}.
The profiles of the $^{12}$CO transitions and HCN(1-0) are all strikingly similar, while the $^{13}$CO(2-1) line profile is narrower.
Our fit yields a broad Gaussian component in the spectrum of HCN(1-0) with $\sigma = 145 \pm 30$ km/s; this value is in agreement with the previously measured $^{12}$CO broad component width of $150\pm7$ km/s \citep{alatalo11a}.
Finally, the peak velocity shifts between the two components are 20 km/s and below for the CO and HCN spectra, and are consistent with zero for all of these spectra except CO(2-1), which has a velocity shift of 10 $\pm$ 4 km/s.

Thus, we interpret this wide component as originating from the molecular outflow, though higher spatial resolution observations are needed to discern the three dimensional kinematics and the outflow.
Hereafter, we refer to the wide component as ``outflow", and the narrow component as ``systemic".
For the $^{13}$CO(2-1), we do not detect the wider outflow component, and are only able to place an upper limit on the flux from the wide component.
We determine this upper limit by simulating spectra with the same Gaussian fit parameters as the observed $^{13}$CO(2-1) spectrum except for the relative amplitudes of the two components.
We add noise to the spectrum consistent with the S/N of the observed spectrum, and find that our non-detection of the outflow component conservatively indicates that the flux of the wide component is no more than 2\% of the narrow component flux, yielding an outflow flux upper limit of $\sim$0.4 Jy km/s.
More detail on this method is given in Appendix~\ref{app:ulim}.

\begin{table*}[]
    \centering
    \begin{tabular}{c|c|c|c|c|c|c|c}
        Line Name & Total Flux & Narrow Flux & Broad Flux & Narrow Peak  & Broad Peak  & Narrow  $\sigma$ & Broad $\sigma$  \\ 
         & (Jy km/s) & (Jy km/s) & (Jy km/s) & Velocity (km/s) & Velocity (km/s) & (km/s) & (km/s) \\ \hline
        HCN(1-0) & 15 $\pm$ 4  & 9.1 $\pm$ 1.4 & 6.1 $\pm$ 3.2 & 2146 $\pm$ 3  & 2159 $\pm$ 17 & 60 $\pm$ 6 & 143 $\pm$ 29  \\
        $^{13}$CO(2-1) [Fit A] & 20.4 $\pm$ 1.1 & 6.2 $\pm$ 0.6 & 14.1 $\pm$ 1.0 & 2143.3 $\pm$ 0.7 & 2158.8 $\pm$ 1.1 & 26.8 $\pm$ 1.1 & 49.1 $\pm$ 0.9 \\
        $^{13}$CO(2-1) [Fit B] & 20.3 $\pm$ 0.4 & 19.3 $\pm$ 0.2 & $\leq$0.04$^{*}$ & 2150.9 $\pm$ 0.2 & 2190 $\pm$ 31$^{\vdag}$ & 39.2 $\pm$ 0.3 & 145 \\
    \end{tabular}
    \caption{Gaussian fitting results for HCN(1-0) and $^{13}$CO(2-1), for the narrow and broad components. For $^{13}$CO(2-1), for fit A we let all parameters vary freely, while for fit B we set the component 2 linewidth to 145 km/s. These fits are plotted with the spectra in Figure~\ref{fig:spectra}. Peak velocities are radio velocities. ($^{*}$) See Appendix~\ref{app:ulim} for how this upper limit was derived. ($^{\vdag}$) this value was at the upper bound of the allowed velocity shifts for this fit.}
    \label{tab:fit_params}
\end{table*}

\subsection{Line Ratios}

\begin{figure}[h]
    \centering
    \includegraphics[width=\linewidth]{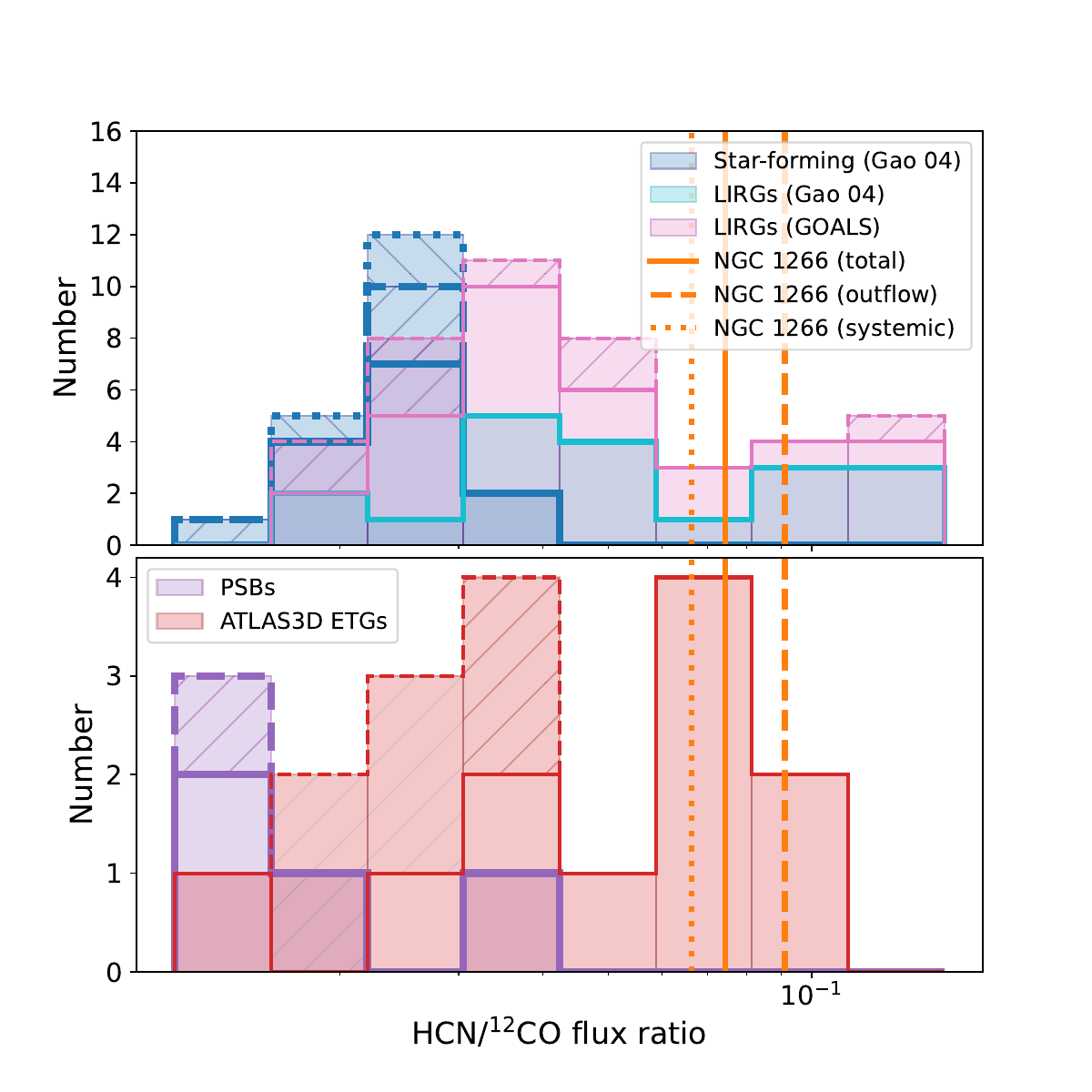}
    \caption{Histogram of the HCN(1-0)/$^{12}$CO(1-0) flux ratio for a variety of galaxy samples. Bars enclosed by solid lines show measured ratios, and bars enclosed by dashed or dotted lines with hatches show upper and lower limits respectively. The solid vertical line corresponds to the measured NGC 1266 value with single dish observations \citep{crocker12}, while the dotted and dashed vertical lines show the decomposed systemic and outflow ratios from this work respectively. The top plot includes star-forming galaxies and LIRGs from \citet{gao04} in blue and teal bars. Pink bars show LIRGs from the GOALS survey \citep{privon15, herrero-illana19}. In the lower plot, red bars show early type galaxies from the ATLAS3D survey \citep{young11}, and purple bars show post-starburst galaxies from \citet{french23a}. The `dense gas fraction' in all components of NGC 1266 is more similar to LIRGs than other early-type galaxies.
    }
    \label{fig:hcn_hist}
\end{figure}

In Figure~\ref{fig:hcn_hist}, we plot the HCN/$^{12}$CO line ratio for the combined, narrow (systemic) , and broad (outflow) components of both lines.
We include integrated measurements of other samples, including star-forming galaxies and LIRGs from \citet{gao04}, LIRGs from the GOALs survey \citep{privon15}, PSBs from \citet{french15a}, and early-type galaxies from \citet{crocker12}.
We see that both the outflow and systemic components of NGC 1266 have relatively high HCN/$^{12}$CO ratios (0.09 $\pm$ 0.06 and 0.066 $\pm$ 0.011, respectively), though some other early-type galaxies and LIRGs host similar ratios.

\begin{figure}[h]
    \centering
    \includegraphics[width=\linewidth]{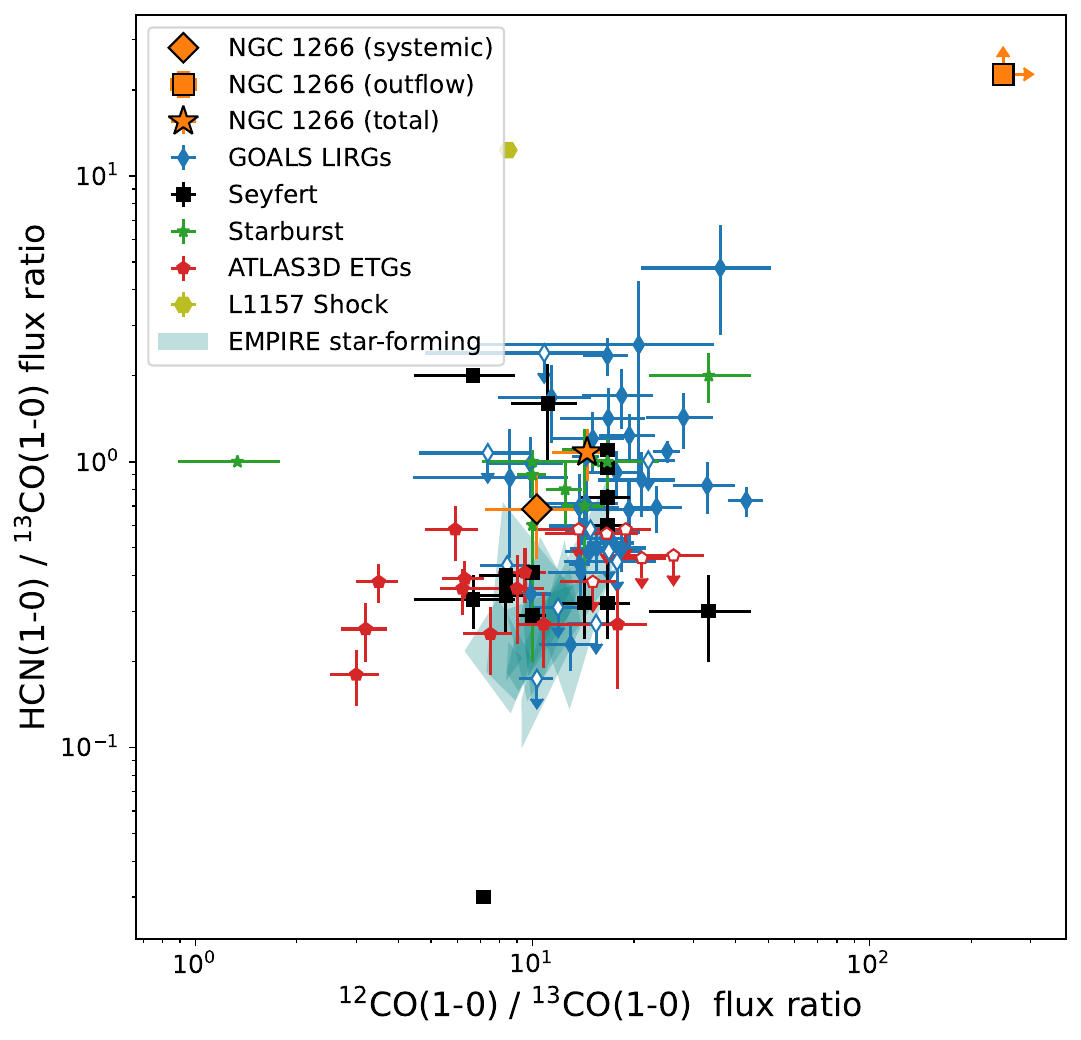}
    \caption{The HCN(1-0)/$^{13}$CO(1-0) flux ratio and the $^{13}$CO(1-0)/$^{12}$CO(1-0) flux ratio for NGC 1266 and a number of other galaxies. The outlined orange hexagon shows the \citet{crocker12} single dish observations, while the outlined orange square and diamond show the decomposed outflow and systemic components from this work respectively. Because we do not detect$^{13}$CO(1-0) in the outflow, the flux ratios for the outflow are limits. In blue diamonds, we plot LIRGs from the GOALS survey \citep{privon15, herrero-illana19}. Black squares show Seyfert host galaxies and green stars show starburst galaxies compiled by \citet{krips10}. Red pentagons show ATLAS3D early type galaxies \citep{young11}. The filled teal quadrilaterals show the range of measured values from spatially resolved observations for each star-forming galaxy in the EMPIRE survey \citep{jimenez-donaire19a}. Finally, the olive hexagon corresponds to the L1157, a shocked region in our Galaxy \citep{yamaguchi12}. Unfilled markers are limits.}
    \label{fig:hcn_ratios}
\end{figure}

In Figure~\ref{fig:hcn_ratios}, we plot HCN/$^{13}$CO(1-0) and $^{12}$CO/$^{13}$CO(1-0) line ratios for NGC 1266 and other samples of galaxies.
We convert our $^{13}$CO(2-1) fluxes and upper limits using the total $^{13}$CO(1-0)/(2-1) ratio of 1.5 $\pm$ 0.4 for NGC 1266 from \citet{crocker12} because the other galaxy samples have $^{13}$CO(1-0) rather than $^{13}$CO(2-1) observations.
In making this conversion, we assume that the integrated $^{13}$CO(1-0)/(2-1) is the same as that of the systemic and outflow components individually.
However, this is an imperfect assumption as we would expect the outflowing gas to be at a higher temperature and thus at higher excitation, resulting in a lower than assumed $^{13}$CO(1-0)/(2-1) ratio and an overestimation of the $^{13}$CO(1-0) flux.
Because we only measure an upper limit for the $^{13}$CO outflow flux, variations in the $^{13}$CO(1-0)/(2-1) ratio make this a more conservative upper limit.

Both plotted line ratios in Figure~\ref{fig:hcn_ratios} vary greatly in the outflow and systemic components.
We measure $^{12}$CO(1-0)/$^{13}$CO(1-0) of 10 $\pm$ 3 in the systemic component and a lower limit of $\geq250$ in the outflow, using the same narrow and broad fit components from the $^{12}$CO(1-0) spectrum \citep{alatalo11a}.
While the systemic component has a similar $^{12}$CO/$^{13}$CO(1-0) ratio to other early-type galaxies, the very large lower limit ratio in the outflow is greater than any of the plotted galaxies.
For the HCN(1-0)/$^{13}$CO(1-0) ratio, the systemic component is consistent with the other early-type galaxies with a ratio of 0.7 $\pm$ 0.2, while the outflow component has another highly elevated ratio of $\leq$23.
We discuss these results in more depth in Section~\ref{sec:discuss-ratios}

We compare our total measured line ratios to the ratios measured with IRAM from \citet{crocker12}.
For HCN/$^{12}$CO, \citet{crocker12} measure a ratio of 0.128 $\pm$ 0.020 for NGC 1266, as compared to the total HCN/$^{12}$CO ratio of 0.075 $\pm$ 0.020 we measure for NGC 1266.
\citet{alatalo11a} compare the $^{12}$CO(1-0) flux they measure with CARMA to the single dish IRAM flux from \citet{crocker12} and find that CARMA recovers 20\% more flux.
To see if a 20\% loss of flux in $^{12}$CO can explain this discrepancy, we divide the $^{12}$CO flux by 1.2, and find an HCN/$^{12}$CO ratio of 0.090 $\pm$ 0.020, nearly consistent with the \citet{crocker12} value within uncertainties.

For the $^{13}$CO line ratios plotted in Figure~\ref{fig:hcn_ratios}, \citet{crocker12} measure HCN/$^{13}$CO(1-0) of 3.94$^{+0.73}_{-0.66}$ and $^{12}$CO/$^{13}$CO(1-0) of 30.9$^{+5.6}_{-5.0}$.
Both of these ratios are greater than our measured values plotted in Figure~\ref{fig:hcn_ratios}, indicating that our observations recover a relatively greater $^{13}$CO flux.
Our greater $^{13}$CO flux measured with ALMA may be due to pointing errors with IRAM; the $^{13}$CO emission is compact (as shown in Section\ref{sec:extent}) compared to the 22\arcsec beam size of IRAM, and thus minor pointing errors could lead to poor flux recovery if the emission is no longer centered in the beam.

\section{Discussion} \label{sec:discuss}

\subsection{Optically thin CO emission in the outflow}

As shown in Figure~\ref{fig:hcn_ratios}, the flux ratio of $^{12}$CO(1-0)/$^{13}$CO(1-0) varies significantly in the systemic and outflow components of the CO emission.
The two factors that typically drive variations in the CO isotopic ratio ($^{12}$CO/$^{13}$CO(1-0)) are varying abundances and the optical depth of the gas. 
The primary abundance effects that may play a role are fractionation, which could increase the $^{13}$CO abundance without requiring a higher $^{13}$C abundance, and a direct enhancement of $^{13}$C from stellar reprocessing \citep{watson76, alatalo15b}.
However, fractionation requires low gas kinetic temperatures (below 35 K) which is unlikely given shock excitation of the molecular gas \citep{pellegrini13a, otter22}.
Stellar reprocessing is also an unlikely driver of the difference in $^{12}$CO/$^{13}$CO(1-0) between the outflow and systemic molecular gas, as the outflowing molecular gas was likely previously intermixed with the systemic gas before being launched by the outflow, less than 10 Myr ago \citep{alatalo11a}.

Thus, the primary difference in $^{12}$CO/$^{13}$CO(1-0) can be attributed to differences in the optical depth of the outflow and systemic molecular gas components.
In a completely optically thin medium, we would expect $^{12}$CO/$^{13}$CO(1-0) to be the same as the $^{12}$CO/$^{13}$CO abundance ratio, which ranges from $\sim$20 to $\sim$100 in the Milky Way, increasing with galactocentric distance \citep[e.g.][]{langer90, sun24, luo24}.
Our observed outflow $^{12}$CO/$^{13}$CO(1-0) lower limit of 250 is far greater than this typical abundance ratio.
This discrepancy indicates that it is likely that the abundance of $^{13}$CO in the outflow is lower than what has been previously measured in the Milky Way.
It is possible that the outflowing molecular gas could include unprocessed gas from outer regions of the galaxy, thus driving down the $^{13}$CO abundance.
However, even with a low $^{13}$CO abundance in the outflow, it is still unlikely that the outflowing molecular gas is optically thick.
High kinetic temperatures and velocity dispersions can decrease the optical depth of a molecular line, and thus a low outflow optical depth is consistent with the measured outflow kinetic temperature of 100 K from \citet{alatalo11a}.

From our measured $^{13}$CO(2-1) outflow flux upper limit we compute an upper limit H$_2$ column density of the outflow component following \citet{pineda10a}, yielding N(H$_2)=7.4\times10^{20}$ cm$^{-2}$, assuming an abundance ratio of $^{12}$CO/$^{13}$CO to be 69 \citep{wilson99} and H$_2$/$^{12}$CO of $10^{-4}$.
Over the circular aperture of the spectrum, we measure an upper limit mass of outflowing CO-emitting gas of 2.8$\times10^7$ M$_\odot$.
This upper limit is consistent with the outflow mass measured by \citet{alatalo11a} of 2.4$\times10^7$ M$_\odot$, using $^{12}$CO observations and assuming the emission is optically thin.

\subsection{HCN diagnostics} \label{sec:discuss-ratios}

The HCN/$^{12}$CO flux ratio is typically considered a tracer for the ``dense gas fraction'' of the molecular gas \citep[e.g.][]{gao04, wu10b, jones23}.
We plot a histogram of the HCN(1-0)/$^{12}$CO(1-0) flux ratio in Figure~\ref{fig:hcn_hist} for the outflow and systemic components of NGC 1266 as well as star-forming galaxies and LIRGs in the top panel, and PSB and early-type galaxies in the lower panel.
The integrated and systemic HCN/$^{12}$CO ratio in NGC 1266 is consistent with typical LIRGs, exceeding the ratio presented by most star-forming and early-type galaxies.
In LIRGs, this high ratio is typically interpreted as indicative of a plethora of cold, dense gas, resulting in enhanced star formation efficiency.
However, in NGC 1266, the overall star formation efficiency is suppressed rather than enhanced \citep{alatalo15c, otter24}, indicating that there is either a significant reservoir of dense gas that is unable to efficiently form stars, or that the HCN emission is enhanced through other means.

However, we note that the HCN/$^{12}$CO ratio is sensitive to varying abundances, excitations, electron densities, and opacities \citep{papadopoulos07, goldsmith17, leroy17}.
In galactic environments and star-forming galaxies, HCN is a relatively reliable tracer of dense molecular gas \citep{wu10b, pety17, onus18}, but in extreme environments like the nuclei of LIRGs, the physical conditions of the gas can be significantly different from typical star-forming environments \citep{davies03, petric11}.
In NGC 1266, the physical conditions of the gas traced by the systemic and outflow components are likely different from each other.
The HCN in the outflow may be enhanced through shocks, as observed in Mrk 231 \citep{aalto12}.
Shocks can increase the HCN abundance due to high temperatures, resulting in more HCN emission, an effect that has previously been observed in galactic outflows \citep{su07, liu11}.
NGC 1266 hosts significant shocked molecular gas in the central 500 pc \citep{otter22}, making shock enhancement of HCN likely.
However, without observations of more HCN transitions, it is unclear how much of the HCN emission is due to shocks and the properties of the HCN-emitting gas.

In addition, despite the high critical density of HCN transitions, gas below this density may dominate the HCN emission with sub-thermally excited HCN.
HCN could be radiatively excited by IR-pumping, or X-ray dominated regions (XDRs) \citep{aalto07, sakamoto10}.
However, previous studies of the high-J CO and H$_2$O spectral line energy distributions exclude significant contributions from either of these mechanisms, instead favoring shock excitation \citep{pellegrini13a}.

\subsection{Dense gas in the outflow}

Overall, the lack of a wide component in the $^{13}$CO line profile is most likely due to the CO emission tracing optically thin gas in the outflow.
However, the broad profile of the HCN emission indicates that there is dense gas also in the outflow.
While these two observations seem contradictory, they can be reconciled by considering a multiphase, clumpy outflow.
In this model, dense clumps of HCN-emitting gas may be entrained in the outflow, resulting in the broad HCN component.
Then the majority of the broad $^{12}$CO emission traces diffuse, optically thin molecular gas surrounding the dense clumps, and no $^{13}$CO is detected in the outflow.
In this scenario, it is unclear whether the HCN-emitting gas was in pre-existing dense clumps of gas that were swept into the outflow, or if the HCN emission originates from molecular gas compressed and heated by shocks driven by the outflow.


The presence of dense molecular gas in the outflow of NGC 1266 makes it one of a handful of systems where dense gas has been detected in the outflow.
These systems include (U)LIRGs \citep[Mrk 231, IRAS 13120-5453]{aalto12, privon17}, starburst galaxies \citep[NGC 253, NGC 1808][]{walter17, salak17}, and Seyfert II galaxies \citep[NGC 1068, M51][]{garcia-burillo14, matsushita15}.
NGC 1266 does not fit any of these categories because it is not actively forming stars beyond the nucleus and lacks a radiatively powerful AGN, but hosts a high surface density, centrally concentrated molecular gas reservoir similar to many of the aforementioned systems.
Further studies of the radio emission of NGC 1266 may reveal whether nascent radio jets are the driving mechanism of the outflow (Otter et. al. in prep.).

\citet{michiyama18a} propose a similar multiphase molecular outflow model for NGC 3256, a dual-nucleus merging LIRG.
The outflow is driven by a low-luminosity AGN with collimated, extended radio structures similar to what we observe in NGC 1266.
In faster regions of the outflow, they observe more HCN emission and lower excitation CO emission, leading them to conclude that the CO and HCN are tracing diffuse and clumpy phases of the outflow, where the high velocity outflow compresses and shocks the molecular gas resulting in excess HCN emission.
Our observations build a similar physical picture, indicating that these properties of molecular outflows may not be unique.

M51, a nearby Seyfert 2 host galaxy also has HCN detected in its outflow, though the proximity of this galaxy enables higher spatial resolution studies.
\citet{matsushita15} resolve multiple clumps of gas in the nuclear region, finding that the kinematics of the CO and HCN emission are similar, including clumps of gas entrained in the outflow.
Therefore, both the dense gas traced by HCN and the more diffuse CO-bright gas are captured by the outflow, as we propose is the model for NGC 1266.

Another object with similar spectral profiles as observed in NGC 1266 is L1157, a galactic dark cloud with ongoing low mass star formation. 
This cloud hosts an outflow with broad profiles in $^{12}$CO, HCN(1-0), and other dense gas tracers, while the line profiles of  $^{13}$CO and C$^{18}$O are narrow.
\citet{umemoto92} conclude that there are dense gas clouds entrained within the molecular outflow, as we observe on a larger scale in NGC 1266.

\subsection{Revised outflow mass}

Though the total outflow mass of NGC 1266 has gone through multiple revisions, with our two phase model and HCN observations we can provide further constraints on the outflow mass.
Initially, the outflow mass was based on the CO(2-1) emission from \citet{alatalo11a}. 
They used RADEX to derive the physical properties of the outflowing molecular gas from the CO(1-0), (2-1), and (3-2) single dish observations, finding a volume density of $n\approx10^3$ cm$^{-3}$ and a kinetic temperature $T\sim100$ K.
Assuming that the CO emission is entirely optically thin, \citet{alatalo11a} derived an outflow molecular gas mass of $2.4 \times 10^7$ M$_\odot$.

However, \citet{alatalo15c} noted that dense gas tracers, including HCN, showed wings in their spectra, thus indicating that the outflowing molecular gas is optically thick.
Using a conversion factor typical for ULIRGs of $\alpha_{CO} \approx 1$ M$_\odot$ (K km/s pc$^{-2}$)$^{-1}$ \citep[e.g.][]{sanders88, downes98}, they measured a molecular outflow mass of $2 \times 10^8$ M$_\odot$, corresponding to a mass outflow rate of 110 M$_\odot$/yr.

To derive an outflow mass from our HCN(1-0) and $^{13}$CO(2-1) observations, we must treat the gas as multi-phase.
The conversion factor from HCN luminosity to molecular gas mass ($\alpha_{HCN}$) is under-studied relative to the more common $\alpha_{CO}$, especially for a galaxy like NGC 1266 with a complex ISM.
For galactic star-forming regions, the typical $\alpha_{HCN}$ employed is $20 M_\odot$ pc$^{-2}$ (K km/s)$^{-1}$ \citep{wu10b}.
However, observations of HCN(1-0) in different conditions indicate that this conversion factor is sensitive to a number of ISM properties, including metallicity, turbulence, excitation properties, and FUV flux \citep{gracia-carpio08, garcia-burillo12, shimajiri17}.
In LIRGs, high amounts of turbulence, temperatures, and excitation in the ISM may all contribute to lowering the value of $\alpha_{HCN}$ \citep{garcia-burillo12}.
Though NGC 1266 is not a LIRG, the shock-excited ISM conditions are likely to be more similar to LIRGs than typical star-forming galaxies, as shown in Figures~\ref{fig:hcn_hist} and \ref{fig:hcn_ratios}, resulting in a likely decreased $\alpha_{HCN}$.
With the great uncertainty in $\alpha_{HCN}$, we calculate the total outflow mass using the canonical value of $\alpha_{HCN} = 20 M_\odot$ pc$^{-2}$ (K km/s)$^{-1}$, resulting in a total outflow mass of 2.2$\times10^8$ M$_\odot$.
Combining this with the diffuse molecular gas, we find a total molecular outflow mass of 2.4$\times 10^8$ M$_\odot$.
To convert this to a mass outflow rate, we assume a similar spatial extent of the HCN(1-0) emission as the $^{12}$CO emission (a maximum outflow extent of 460 pc) and thus use the same dynamical time of the outflow as \citet{alatalo11a} of a conservative 2.6 Myr.
These assumptions yield a total mass outflow rate of 85 M$_\odot$/yr.
We emphasize that this outflow rate is valid only if the HCN-emitting gas in the outflow is in relatively typical star-forming conditions, so it should instead be considered an upper limit on the total outflow mass because $\alpha_{HCN}$ is likely overestimated.
This upper limit is below the \citet{alatalo15b} outflow mass rate of 110 M$_\odot$/yr because they assume the CO emission in the outflow is optically thick.
While the lack of a broad component in the $^{13}$CO(2-1) spectrum indicates that the CO emission is actually optically thin, our mass outflow rate assumes the HCN(1-0) is optically thick; which is presumably a more robust assumption because HCN traces higher density gas, and is thus more likely to be optically thick.
To verify the optical depth of the HCN emission, observations of isotopologues are necessary, while observations of higher-J HCN emission lines would constrain the excitation properties of the HCN.
While our observations include a detection of H$^{13}$CN(3-2) and HC$^{15}$N(3-2) (fluxes reported in Appendix~\ref{app:lines}), the spectra are too noisy to identify whether a broad, outflow component is present.

The large mass outflow rate we compute is surprising given the likely low bolometric luminosity of NGC 1266's AGN.
From \citet{fiore17}, other galaxies with molecular outflows of similar mass outflow rates ($20\sim100$ M$_\odot$/yr) tend to have AGN with bolometric luminosities ranging from 10$^{44}$-10$^{45}$ erg/s, multiple orders of magnitude greater than previous estimates of NGC 1266's bolometric luminosity of 10$^{42}$-10$^{43}$ erg/s \citep{alatalo15c, chen23}.

\subsection{NGC 1266's outflow in context}

Our measured HCN(1-0) outflow mass rate bolsters previous claims that the outflow in NGC 1266 is AGN-driven rather than driven by residual nuclear star formation.
Though measurements of the star formation rate in NGC 1266 vary, the highest upper-limit star formation rate estimate is from the total infrared luminosity, yielding a value of 2.2 M$_\odot$/yr, far too low to power a mass outflow rate of 85 M$_\odot$/yr \citep{alatalo15c}.
Further, NGC 1266 likely hosts a weak AGN, as shown in spectral energy distribution fitting \citep{alatalo15c, chen23}, and observed directly with the Very Long Baseline Interferometer \citep{nyland13}.

However, the role of this likely AGN-driven outflow in the evolution of NGC 1266 is still unclear.
While AGN-driven outflows are often cited as a driver of gas expulsion during quenching, many observed outflows in the nearby universe appear unable to effectively eject the molecular gas reservoirs \citep[e.g.][]{luo22}.
For galaxies with non-ejective outflows, it is unclear whether the outflow contributes to star formation regulation by driving turbulence through the gas reservoirs, or if these outflows are concurrent with quenching but not physically necessary for quenching to occur.

In NGC 1266, we consider whether the observed outflow is capable of expelling the molecular gas reservoir.
As noted in \citet{alatalo15c}, only a fraction of the molecular gas in the broad component exceeds the escape velocity and thus could be ejected from the galaxy.
As we measure the same broad component velocity width in the HCN(1-0) emission as \citet{alatalo11a}, using their derived inner-disk escape velocity of $v_{esc}=340$ km/s, only $\sim$2\% of the molecular gas in the outflow may be ejected.
Then, the outflow mass \textit{escape} rates are only $\leq$0.2 M$_\odot$/yr and 1.7 M$_\odot$/yr for the $^{13}$CO(2-1) and HCN(1-0) respectively.
With a molecular gas reservoir of 1.1$\times10^9$ M$_\odot$ \citep{alatalo11a}, if the outflow were to continue at this mass ejection rate, the depletion time for the molecular gas reservoir would be $\sim$650 Myr.
If we include star formation as a source of gas depletion, using the nuclear star formation rate of 0.7 M$_\sun$/yr \citep{otter24}, the gas depletion time is reduced to 450 Myr. 

Given that NGC 1266 underwent a quenching event $\sim$500 Myr ago \citep{alatalo14b}, if star formation and the outflow were to continue removing molecular gas at this rate, NGC 1266 would maintain a reservoir of molecular gas for about 1 Gyr after beginning to quench.
This timescale is consistent with the finding of \citet{french18} that post-starbursts tend to lose their molecular gas approximately 0.7 - 1.5 Gyr after the peak of the starburst.
While we are unable to predict whether the outflow in NGC 1266 will continue ejecting molecular gas at a similar rate, our study of NGC 1266 indicates that gas removal through the post-starburst phase could be plausibly achieved through relatively low-level AGN activity and residual star formation.

However, the outflow may still have impacts on star formation regulation on shorter timescales before the gas is expelled.
Previous works have proposed that the outflow in NGC 1266 injects excess turbulence into the surrounding ISM, thus suppressing star formation \citep{alatalo11a}.
Turbulent star formation suppression due to AGN feedback is an attractive mechanism by which post-starburst galaxies are able to retain their molecular gas reservoirs after quenching their star formation, and a few PSBs have been observed to have potentially high turbulent pressures in their ISM \citep{smercina22b}.
While very high spatial resolution molecular line observations are necessary to study this turbulence directly, the presence of HCN(1-0) emission in the molecular outflow indicates that denser phases of molecular gas are swept in the outflow, though the origin of this gas is still unclear.
Higher resolution, multi-transition observations are necessary to determine whether the outflow heats the gas kinetically (i.e. by driving turbulence), through shock excitation, or a combination of these two.
In total, our work is consistent with the outflow playing a twofold role in star formation regulation after the quenching episode in NGC 1266: first by disrupting dense gas necessary for star formation, and secondly by slowly expelling the remaining molecular gas, ensuring that the galaxy remains quiescent (barring further cold gas accretion).

\section{Conclusions} \label{sec:conclusions}

In this work, we present ALMA Cycle 0 observations of HCN(1-0) and $^{13}$CO(2-1) in NGC 1266.
We make the following conclusions:
\begin{enumerate}
    \item We detect a broad kinematic component in the HCN(1-0) spectrum with very similar kinematics to the molecular outflow previously detected in $^{12}$CO, and conclude that there is HCN-emitting gas in the outflow.
    \item We do not detect any broad kinematic component in the $^{13}$CO(2-1) emission. We conclude that the CO-emitting gas in the outflow is optically thin.
    \item The systemic component of the emission has a $^{12}$CO/$^{13}$CO and HCN/$^{13}$CO flux ratios consistent with other early-type galaxies, whereas the lack of a $^{13}$CO detection in the outflow drives high $^{12}$CO/$^{13}$CO and HCN/$^{13}$CO flux ratios.
    \item The HCN-emitting gas likely traces a denser phase of the outflow than the diffuse gas traced by optically thin CO emission.
    This denser phase may be concentrated in clumps entrained in the outflow, while the surrounding molecular gas traced by CO is warm, diffuse, and optically thin. We are unable to determine whether the HCN-emitting gas was previously in clouds that were swept into the outflow, or if the high velocity of the outflow shocks and compresses molecular gas, making the molecular gas more dense.
    \item We measure a revised upper limit outflow \textit{escape} rate of 1.7 M$_\odot$/yr by assuming a standard $\alpha_{HCN}$ of 20 M$_\odot$ (K km/s)$^{-1}$. This value yields a minimum molecular gas depletion time of 450 Myr.
    \item This depletion time indicates that low-level AGN activity is likely capable of slowly expelling the molecular gas through the post-starburst phase, potentially resulting in long-term quiescence.
\end{enumerate}

High spatial resolution HCN observations are needed to confirm the multiphase nature of the molecular outflow by directly observing entrained clumps of dense gas in the outflow.
Further, a variety of HCN transitions would also allow greater characterization of the entrained clouds: it is unclear how dense or warm these clouds are, and how much of the HCN emission is due to shock excitation. 
Finally, more detailed study of outflow driving mechanisms, such as potential nascent radio jets in NGC 1266, are needed to determine whether the current outflow in NGC 1266 is likely to continue as these jets grow, or if the outflow is a remnant of previous AGN activity that is unlikely to continue at a consistent rate.
Overall, our results indicate that AGN feedback could plausibly regulate star formation in multiple ways: on short timescales by heating dense molecular gas thus suppressing ongoing star formation, and over $\sim$1 Gyr timescales by slowly expelling the remaining molecular gas reservoirs, leaving the galaxy quenched and devoid of gas.

\section{Acknowledgments}

The authors would like to thank the anonymous referee for their constructive feedback.

J.~A.~O.~acknowledges support from the Space Telescope Science Institute Director's Discretionary Research Fund grant D0101.90311 and funding from the Maryland Space Grant Consortium.
P.P. and M.S. gratefully acknowledge support from the NASA Astrophysics Data Analysis Program (ADAP) under grant 80NSSC23K0495.
Y.L. acknowledges support from the Space Telescope Science Institute Director's Discretionary Research Fund grant D0101.90281.
A.M.M. acknowledges support from the NASA Astrophysics Data Analysis Program (ADAP) grant number 80NSSC23K0750, from NSF AAG grant \#2009416 and NSF CAREER grant \#2239807, and from the Research Corporation for Science Advancement (RCSA) through the Cottrell Scholars Award CS-CSA-2024-092.

This paper makes use of the following ALMA data: ADS/JAO.ALMA\#2011.00511.S. ALMA is a partnership of ESO (representing its member states), NSF (USA) and NINS (Japan), together with NRC (Canada), NSTC and ASIAA (Taiwan), and KASI (Republic of Korea), in cooperation with the Republic of Chile. The Joint ALMA Observatory is operated by ESO, AUI/NRAO and NAOJ.
The National Radio Astronomy Observatory is a facility of the National Science Foundation operated under cooperative agreement by Associated Universities, Inc.

\facilities{ALMA}
\software{numpy,matplotlib,astropy,spectres,casa,scipy}

\appendix

\section{Molecular Line Detections} \label{app:lines}

In addition to the molecular lines discussed in the main text, the observations in Program 2011.0.00511.S include detections of a number of other molecular lines.
We present a list of these molecular lines and their measured fluxes in Table~\ref{tab:all_lines}.
To improve our signal to noise ratio, we spectrally rebin each spectrum using \texttt{SpectRes} \citep{carnall17a} to a spectral resolution of 33 km/s, and fit each line with a Gaussian as in Section~\ref{sec:fitting}. 

However, we are unable to accurately determine the continuum levels of the emission due to the lack of continuum-only spectral windows and high density of spectral features in some spectral regions.
Due to the lack of reliable continuum measurements, we opt to fit a flat continuum for each spectrum by selecting line-free channels.
We fit a flat continuum, and then estimate a range of possible continuum values by-eye.
We then fit the lines using these different continuum values and use the maximum and minimum fitted fluxes to estimate the uncertainty.

\begin{table}[]
    \centering
    \begin{tabular}{c|c|c}
     Emission Line & Rest frequency (GHz) & Flux (Jy km/s) \\ \hline
     H$^{13}$CO+(4-3)$^{(a)}$ & 346.998344 & 3.5$^{+0.3}_{-1.4}$ \\
     SiO(8-7)$^{(a)}$ & 347.330631 & 0.25$^{+0.81}_{-0.01}$ \\
     SiO(7-6)$^{(b)}$ & 303.92696 & 0.7$^{+2.2}_{-0.01}$ \\
     CH$_3$OH 2(1,1)-2(0,2)$+-$$^{(b)}$ & 304.208 & 5.0$^{0.4}_{-0.9}$ \\
     CH$_3$OH 1(1,0)-1(0,1)$+-$ & 303.367 & 3.0$^{+0.5}_{-0.2}$ \\
     H$^{13}$CO+(3-2)$^{(c)}$ & 260.255339 & 1.00$^{+0.8}_{-0.08}$ \\
     SiO(6-5)$^{(c)}$ & 260.51802 & 1.1$^{0.4}_{0.13}$ \\
     HN$^{13}$C(3-2) & 261.263 & 0.97$^{+0.5}_{0.07}$ \\
     HC$^{15}$N(3-2) & 258.157 & 1.09$^{+1.11}_{-0.7}$ \\
     H$^{13}$CN(3-2) & 259.012 & 2.0$^{+1.0}_{-0.3}$ \\
     SiO(5-4) & 217.10498 & 1.11$^{+0.6}_{-0.2}$ \\
     H$_2$S 2(2,0)-2(1,1) & 216.71 & 0.44$^{+0.7}_{-0.02}$ \\
     SiO(2-1) & 86.84696 & 0.96$^{+0.2}_{-0.02}$
    \end{tabular}
    \caption{Measured fluxes for all molecular lines detected. Lines with matching superscripts are blended, and may be contaminated.}
    \label{tab:all_lines}
\end{table}

\section{Two component upper limit estimation} \label{app:ulim}

To estimate an upper limit on the relative flux of a non-detected, broad ($\sigma$ = 145 km/s) Gaussian component in the $^{13}$CO(2-1) spectrum shown in Figure~\ref{fig:spectra}, we use the following method.
We simulate the $^{13}$CO(2-1) spectrum with two Gaussian components and add noise such that the peak signal to noise ratio is 88, as we measure in the observed spectrum.
We measure the SNR in the original spectrum by taking the peak flux divided by the standard deviation of the line-free continuum regions.
We set the widths of both Gaussian components equal to the widths derived from the constrained fit of $^{13}$CO(2-1) (Fit B) in Table~\ref{tab:fit_params}.
The width of the broad component, $\sigma$ = 145 km/s was held constant for this fit, as this is the width of the broad component in the HCN(1-0) and $^{12}$CO spectra.
We create 100 simulated spectra, with the ratio of amplitudes between the two Gaussian components ranging from 1e-4 to 1.
We then fit each simulated spectrum with two Gaussian components, using the same methodology used to fit our observed spectrum, and measure our recovered amplitude ratio between the two fitted components.

\begin{figure}
    \centering
    \includegraphics[width=0.5\linewidth]{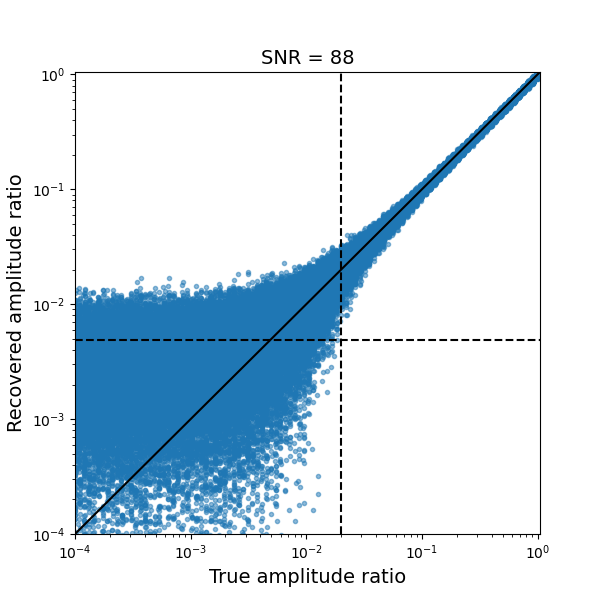}
    \caption{We plot the true amplitude ratio of a simulated spectrum of a narrow and wide Gaussian with noise added versus the amplitude ratio recovered by our fit in blue points. The solid line is where the true and recovered amplitude ratios are the same. The horizontal dashed line corresponds to the recovered amplitude ratio of our observed $^{13}$CO(2-1) spectrum, and the vertical dashed line is the upper limit true amplitude ratio for this value.}
    \label{fig:ampratio}
\end{figure}

We repeat this process 1,000 times, and plot the recovered amplitude ratio versus the true amplitude ratio in Figure~\ref{fig:ampratio}.
We plot a dashed horizontal line at our measured amplitude ratio from the observed spectrum of 0.005.
The maximum true amplitude ratio corresponding to the this recovered amplitude ratio in our simulations is about 0.02.
Hence we conservatively adopt 0.02 as an upper limit amplitude ratio between the two components in our observed spectrum, indicating that the amplitude of the broad component in our observed $^{13}$CO(2-1) spectrum is at most 2\% of the amplitude of the narrow component.

\bibliography{mol_lines_ngc1266}{}
\bibliographystyle{aasjournal}

\end{document}